Preprint /2020

# Shadow economy and populism – risk and uncertainty factors for establishing low-carbon economy of Balkan countries (case study for Bulgaria)

**Shteryo Nozharov[a], Nina Nikolova[b*]**

[a]*Department of Economics, University of National and World Economy, 1700 Sofia, Student Town, Bulgaria*
[b]*Department of Climatology, Hydrology and Geomorphology, Sofia University "St. Kliment Ohridski", 1504 Sofia, 15 Tzar Osvoboditel Blvd., Bulgaria*

***Abstract.*** *The main purpose of the current publication is to formulate a scenario model for analysis of the opportunities for low-carbon economy establishment in the countries with transition economies. The model studies risk factors such as shadow economy level and populism based on the implementation and development of Balkan countries' economic policy and at the same time shows future climate changes tendencies and uncertainties of climate models. A transdisciplinary approach is implemented in the study. Climate change perception and understanding about low-carbon economy are examined through the public opinion and analysis of mass-media publications. The results of the research are important in order to clarify the multicultural divergences as a factor for risk and uncertainty in the implementation process of the policy for climate change. In this way geographical aspects of risk and uncertainty, which are not only related to the economic development of the relevant countries, could be brought out.*

[*]Corresponding author: E-mail address: nozharov@unwe.bg (S. Nozharov)

# 1 Introduction

Climate change is one of the most significant challenges in the contemporary world which has impact on various sectors of the economy. That is why the interest to this topic has been increased in many research works. The scientific publications show increasing of air temperature during last decades. Hartmann et al. (2013) show that the first decade of 21-st century is the warmest period during the instrumental observations. According to WMO (2020) the past five years and the past decade were the warmest on the record. The global mean temperature in 2019 was about 1.1 °C above pre-industrial levels (1850–1900). For the last 50 years fossil $CO_2$ emissions represent 82 % of the total emissions in global carbon budget (WMO, 2020).

The main purpose of the current study is to contribute to the risk and uncertainty forecasting models in the context of the climate change policy. In this regard the transdisciplinary approach is applied. This allows creation of a conceptual model of the research problem in order to integrate in a general perspective the climatic and economic analyses. In this way the specific restrictions of the separate disciplines are eliminated. In the present paper the vision of a low-carbon economy goes beyond traditional macroeconomic models as the economic system is analyzed as part of the global biosphere. On the other hand the climatic analysis overcomes the restrictions of the ecological approach as the climate processes are analyzed in the conjunction with the economic activity dynamics.

The present paper makes connection among the risk and uncertainty of the climate change policy, polluting infrastructure of the Bulgarian energy system and the multicultural differences. These issues are examined through the level of shadow economy and populism.

In order to achieve the aim of the study a transdisciplinary approach has been implemented which includes qualitative and quantitative methods. The qualitative analysis explores through a questionnaires how Bulgarian society perceives the climate change. An internet-based inquiry of the Bulgarian mass media about the presence of populism in the coverage of news related to climate change has also been carried out. The online surveys was used to examine the extent to which the society understands the problems related to climate change and the perception of measures to limit greenhouse gas emissions into the atmosphere. The quantitative analysis is based on the climate and economic data. The paper gives the information about the expected changes in air temperature in the study area as well as shows the degree of influence of the shadow economy and populism on the achievement of a low carbon economy in Bulgaria. The data from International Monetary Fund databases and Eurostat is used for the economic analysis.

The results of the present study could complement the concept of Bridge et al. (2013) for the Geography of Energy Transition: Space, place and the low-carbon economy in terms of how Bulgaria and the other Balkan countries would fit into such model. Is it possible to identify the risk and uncertainty in reaching a low-carbon economy in a geographical perspective?

Next, analysing the specificities of the countries with economies in transition, the study could contribute to the analysis of multicultural differences identified by IPCC as a factor of risk and uncertainty in climate change policy (Kunreuther et al., 2014).

The current study describes also the climate change tendencies in South-east Europe which are characterized by the risk of drought on the one hand and the extreme precipitation and floods on the other. The analysis of precipitation variability during the 20-th and in the beginning of the 21-st century, as well as the results from climate models show the general decreasing trend in southern Europe, including Bulgaria (IPCC 2007, EEA, 2012, PRUDENCE, 2005). Chenkova and Nikolova (2015) point out the increasing of air temperature and decreasing of precipitation in Bulgaria in the future (2051-2080) which could increase the risk of drought events. Despite the general tendencies to drought the increasing of the frequency of extreme precipitation was observed in the recent years. This causes significant floods and creates a risk for the population. Bocheva et al. (2010) found that during the period 1991-2007 the average number of days with daily rainfall values over 100 mm increased by about 30% compared to the baseline period 1961-1990.

The paper has the following structure: 1. Introduction, 2. Theoretical background, 3. Study area data and methods, 4. Theory / calculation, 5. Results and discussion and 6. Conclusion. The second section of the paper analyses main definitions related to risk and uncertainty in the climate change policy as well as makes the review of existing research on the investigated topic. The data, qualitative and quantitative methods used in the present study are presented in the third section. On the background of climate change risk assessment the fourth part of the paper reveals the uncertainty of climate models and the necessity of observation data for climate analyses. The results and discussions are related to the investigation about public perception about climate change and related issues as well as present in-depth statistical and economical analysis of the relationship shadow economy – populism – climate change policy.

# 2 Theoretical background

## 2.1 Risk and uncertainty definition in the context of climate change analysis

After the studies of Wunne (1992) and Stirling (2007) were published, the definition of risk and uncertainty in the context of environmental policies has a new meaning. Four different categories were distinguished: risk, uncertainty, ignorance and indeterminacy. Kunreuther et al. (2014) specifies that one of the gaps in the skills and knowledge are related to the multicultural differences in regards to human perception and behavior to the climate change. The current study analyzes the multicultural differences of the Balkan countries with the economies in transition.

The research question is if the multicultural differences of Balkan countries with transition economies could be accepted as an independent factor which impacts over the risk and uncertainty of the climate change policy. The research hypothesis is that the shadow economy and the populism are multicultural differences, which are typical for the Balkan countries and they impact on the risk and uncertainty of the climate change policy.

The analysis of the share of shadow economy as a percentage of the GDP for 31 European countries, according to a study of the International Monetary Fund (Medina and Schneider, 2018), indicates that the first four countries in the ranking are transition economy countries from the Balkans (Bulgaria, Turkey, Croatia and Romania). The share of the shadow economy as a percentage of GDP in these countries is 25-30% for 2017 while in Switzerland and Germany this percentage is 6-10%. The main purpose of the shadow economy is to reach a profit at any cost while climate change and compliance with environmental standards are not a priority. These conclusions are also confirmed in publications, written by Bulgarian authors (Stoykov and Dimitrova, 1999; Vassileva, 2007).

There are also favorable conditions for the development of populism. That is why, the uncertainty in the behavior of some economic agents in transition economy countries (which are also EU member-states) could cause risk to the global efforts for climate change. The multicultural differences of these countries, compared to the other EU member-states are based on the fact that these countries are former socialist countries. The informal public and social networks in these countries are still aware of the concept for totalitarianism, that the resources are cheap and unlimited, that the smoking factory's chimneys are symbol of progress because they ensure employment for the population (Lavigne, 1999; Kronenberg, 2004). These social networks impact over the government policy as they are the main electorate. For example, Bulgaria has been an EU member-state for ten years but many electric power stations which operate on lignite and pollute the environment have not been closed so far.

The effects over risk and uncertainty could be broader. Three of the Balkan countries are EU member-states (Bulgaria, Romania and Croatia). The non-execution of their national priorities for low carbon emissions could impact over the EU policy for climate change.

In the literature review are identified the papers of Balintet et al. (2017), Brunner et al. (2012), Farmer et al. (2015), Peterson, (2006). They examined these issues in the absence of political will and political instability. From this point of view, the present study is the first of its kind puts the accent on the multicultural differences, shadow economy and populism.

## 2.2 The relationship shadow economy and populism – risk and uncertainty in the climate change policy

In 2016, the World Health Organization (WHO, 2016) ranked Bulgaria on a second place in the world ranking because of the polluted air and the number of respiratory diseases caused. In 2017 the EU Court of Justice judged Bulgaria for the same reason (Case C-488/15). Consequently, a climate imaginary in the type "fossil fuels forever" is established in Bulgaria. The Bulgarian energy sector is low efficient, dependent on the lignite and highly polluting the air and the environment.

According to Dimitrova (2011), the Bulgarian energy sector is a black box. In the energy system public information about the cash and material flows is missing. The same statement is also approved by the Center for the Study of Democracy (2015). They consider that the Bulgarian energy sector is not governed transparently. The main energy resource is the highly-emission lignite which produces 37% of the gross domestic energy consumption in Bulgaria. Another problem is the high transformation, transportation and consumption losses of energy in Bulgaria, which are almost 50%. In general, the transformation and transportation losses of energy are equal to the volume of energy produced by renewable energy resources. European Commission (COMP/39.767-2015) states that the energy market liberalization in Bulgaria is too slow and formal.

The empirical testing of Borlea et al., (2017) shows the relationship among corruption, shadow economy and economic growth for the EU member-states. According to this study for the period 2003-2014 the Balkan countries Bulgaria and Romania have the greatest share of the shadow economy and corruption, compared to the other EU countries and the lowest living standard level (measured by GDP per capita). The authors make the conclusion that there is a relationship between the aforementioned indicators and the Bulgaria and Romania affiliation to the former communist bloc. It is also indicated that the transition economy countries, which are former communist countries have multicultural differences with the other EU member-states. A similar relationship between shadow economy and post-socialist type of countries has been established by the investigation of Bayar and Ozturk (2016) which examine the financial development and shadow economy in the European Union transition economies. The results of the study show that Bulgaria, Romania and Croatia have the greatest share of shadow economy for the period 2003-2014.

According to Stavrev (2017), in the last decade, the Bulgarian government has restricted the control over the monopolies and cartels, including the observation of the environment protection standards. The society has no access to the information, concerning the main priorities of the government policy. This results in missing values of the society, which is a good basis for development of populism. This hypothesis is confirmed by other Bulgarian authors. Badjakov (2013) examines the suspicions about Non-public Politics (Backstage Politics) in the Bulgarian society. This means that people do not worry about the issues, related to the ecological problems and climate change. The everyday problems of the society are related to the conspiratorial theories about the underhand governance of the state. This also creates favorable environment for development of the populism.

The classical economic definition of the term populism is presented in the publication of Dornbusch and Edwards (1990) concerning the macroeconomics of populism. The term means political mobilization with repeating rhetoric and symbols whose main purpose is a huge number of people to be inspired for a reform of the economic policy. In the context of the climate change policy, there is no clear definition of the term climate populism. There are many viewpoints and concepts. For example, Levy and Spicer (2013) use the term "climate imaginaries" according to which they define the shared socio-semiotic systems which generate group understanding of the climate change. Four types of the climate imaginaries are pointed out in the cited publication: "fossil fuels forever", "climate apocalypse", "techno-market", and "sustainable lifestyle". Swyngedouw (2010) uses the term

„apocalyptic imaginaries“ in the similar sense in order to introduce the climate populism to the policy makers. Another term which is used in relation to the climate populism is the discourse. It is introduced for the first time by McDonald (2015) and describes the public marginalization of the discourse for climate and ecologic security against the discourse for energy and economic security. Dunlap (2013) presents the term „Climate Change Scepticism“ which covers the cases when the existence of the problem is rejected. In this way, the term „Climate Change Scepticism“ is close in definition to the climate populism and it is a part of it.

**Based on the literature review and for the needs of the research, we have elaborated the following definition of the term „Climate Change Populism”: It is a repeated marginal rhetoric addressed to a wide range of public groups which main purpose is to legitimate a definite climate imaginary, in favor of separate business groups, which fact will destroy the public interest in balanced sustainable development. This can either force the government to take excessive measures in order to approach low carbon economy or to completely reject such measures.**

There are open channels for marginal climate rhetoric in Bulgaria. According to Stavrev (2017), the following is observed about the Bulgarian society in 2017: NGOs which defend exotic thesis and media, whose ownership and resources of funding is unknown and they work at a loss. The same statement is approved by the Center for the Study of Democracy (2016). According to the center, for the period 2005-2014 most of Bulgarian print media work at a loss and their purpose is to exert media impact. The ownership of part of these media is unclear and 70% of this ownership is divided between two owners who develop the economic activities in other sectors. Consequently, there are open media channels and suspicious NGOs, which propagate marginal climate rhetoric.

## 3 Study area, data and methods

The scope of the research are Balkan countries with the economies in transition and Bulgaria in particular. The transition economy countries have socio-cultural and economic similarities, which distinguish them from the other two categories of countries (Lavigne, 1999; Kronenberg, 2004). Most of these countries have been Former Soviet Union members and they are industrial countries. Moreover, these countries accept the EU climate change policy, and some of them are EU member-states, EU Candidate Countries or EU border area. In this way, the transition economy countries are much alike the developed countries. On the other hand, in these countries the GDP per capita is too low, the share of the shadow economy is too high as well as the level of corruption. According to the aforementioned, the transition economy countries are much alike the developing countries. From socio-cultural perspective, the risk and uncertainty, typical for the developed and developing countries, could not be the same as the risk and uncertainty in transition economy countries. That is why these countries are chosen as a subject of the current research. The accent of the analysis is the situation in Bulgaria.

According to the International Monetary Fund (Murgasova et al., 2015) Bulgaria, together with Romania and Croatia are included in the Emerging Europe group where they are defined as transition economy countries. These countries are distinguished from the Advanced European Union countries (EU 17).

The following type of data and information are used for the purpose of present analysis:

1. Information from mass media. In order to examine the populism in climate change topic an inquiry of the mass media publications in Bulgaria for the period 2012-2017 is done. Five of the most popular private electronic media in Bulgaria are chosen for the purpose of the analysis (https://www.capital.bg; https://www.investor.bg; https://www.24chasa.bg;https://trud.bg;https://www.mediapool.bg/). 500 publications on the topic for climate change are studied. The expert assessment method is applied in order to be analyzed each of the publications.

2. Answers of the questionnaire distributed among the people representing various sectors in Bulgaria. The main topic of this questionnaire is climate change and low carbon economy.
3. Climate data: model data about air temperature from MPI-EMS-MR, Max-Plank Institute, Germany are used for the RCPs (Representative Concentration Pathways) 4.5 and 8.5. The data are available on the web site of Nederland Meteorological Institute (https://climexp.knmi.nl/start.cgi?id=someone@somewhere, accessed by April 2020).
4. Data for the economic analysis: Greenhouse gas emissions intensity of energy consumption (Dataset, Eurostat, 2020) as an indicator for national climate change policy; data for shadow economy (International Monetary Fund databases, Medina and Schneider, 2018). Data for Shadow economy from the IMF database are selected as the most-up-to-date and reliable ones. There are also publications of Bulgarian authors, which are published in world's largest and indexed databases and consist of reliable data for shadow economy. Unfortunately the presented information about shadow economy in these publications is reliable one but it is not up-to-date (Stoykov and Dimitrova, 1999; Vassileva, 2007).

## Quantitative methods

The results of the present study are obtained by quantitative and qualitative analyses. The main purpose of the quantitative analysis is to integrate in a general perspective the climatological and economic analyses. The climate change impact on the shadow economy level and populism over the low carbon economy in Bulgaria are examined. As a quantitative method the regression analysis is applied on two aspects: 1) to determine the tendencies in seasonal air temperature changesin the investigated area for the 2021-2050 and 2051-2080 periods and 2) to analyze the shadow economy level and the climate populism impact on the national climate change policy in Bulgaria.

The tendencies in multi-year course of air temperature is analysed by application of linear regression model ($y=b_0+b_1*x$) of the time-series. The linear regression is a widely used method in the specialized scientific literature on climatology to study trends in the multiyear course of climatic parameters (Easterling et al., 1997; Smith, 2008; Wang, 2009; IPCC, 2007, 2013; Peng-Fei et al., 2015; Chattopadhyay and Edwards, 2016). The calculation of the trend and evaluation of its statistical significance by T-statistics at level 0.05 were done in AnClim software (Stepanek, 2008).

The regression method was used in order evaluate combined impact of the factors on national climate change policy. For the economic assessment the regression analysis is based on the dependent variable “Greenhouse gas emissions intensity of energy consumption” and the independent variables shadow economy and climate populism. The data for Greenhouse gas emissions intensity of energy consumption characterize the national climate change policy.

The model of Hammonds et al., (1994) is used in order to measure the risk and uncertainty in the accomplishment of the main purposes of the Bulgarian policy on climate change till 2030. The information is taken from the official website of the Bulgarian Ministry of Environment and Water (MOEW, 2012).

## Qualitative methods

The qualitative analysis is conducted in order to analyze the public opinion about climate change, climate change policy and low carbon economy. The present study aims to give the answer of the question “How the society in Bulgaria perceive the issue of "climate change" and what is their understanding of setting targets and implementing measures to limit greenhouse gas emissions in the atmosphere?” For the purpose of the present paper the questionnaire of 12 questions was cerated and was distributed by e-mail to the people representing different sectors and categories in Bulgaria such as industry, services, agriculture, administration (including state administration), science, non-governmental organizations, and students. The questionnaire was made through a free survey form at https://www.surveyrock.com and was made also available to the wide public by announcement on social media which gave the possibility for participation for the people not listed above.The number of questions is limited to 12 in order to use the possibilities for a free survey, but in our oppinion the information from the answers of these questions is sufficient to present the general picture in Bulgaria about climate change perception and the transition to low-carbon economy issue. The qualitative information from the surveys was evaluated by the relative share of the answers to the individual questions (in %) and the results are presented graphically.

Apart from the introductory question on the affiliation of respondents to the different sectors, the other questions of the survey give information that can be referred to the following two groups: 1. How does the population understand / accept the problem of "climate change" and 2. What does Bulgaria do about reducing the greenhouse gas emissions into the atmosphere, which stimulates the transition to a low-carbon economy in the country and which are the obstacles for this?

The second aspect of the application of the qualitative analysis is for the examination of the existence of populism on the topic of climate change in Bulgaria mass media for a period 2012-2017. For the purposes of the analysis five of the most popular private electronic media were examined and 500 publications on the topic of climate change were analyzed through the expert assessment method.

In order to reveal whether the presence of populist media publications on climate change influences public opinion the comparison between the analyses of public opinion and mass media publications was made.

## 4 Theory/calculation

The theoretical knowledge, observations, and climate models data demonstrate that there is a link between climate change and the growing risk of the following climatic extremes: extreme heat, heavy rainfalls (including snowfalls), drought and forest fires (Huber and Gulledge, 2011). The holistic approach for climate change risk assessment includes the following steps: 1) analysis of future pathway of global greenhouse gas emissions; 2) reaviling the physical climate, and their direct risks to human activity; 3) assessment of the risks arising from interactions between changes in the physical climate and human systems, and 4) valuing the changes that might take place (King et al., 2015).

One of the main tasks for the scientific community dealing with modeling climate change is the assessment of the uncertainty of the climate models (Bader et al, 2008). The accuracy and details of the experiment depend most on horizontal resolution of the models. The reduction of the step is accompanied by the need to significantly increase the computing resources. Model results should not be considered as a forecast of the weather in a future period. Simulations of climate change should be considered as a comparison of the average values of the model variables (temperature, precipitation, etc.) in the past and the future. Typically, 30-year periods are compared.

The IPCC (2013) develops the RCPs (Representative Concentration Pathways) scenarios which show the radiative forcing due to the increased concentration of greenhouse gases. Depending on the population, gross domestic product, technology development, food and electricity production, electricity consumption and land use change, the scenarios give different concentration trajectories. According to the RCPs scenarios the implementation of measures in the area of climate change could lead to a stabilizing effect and even to a reduction in the radiative forcing, which will also affect the strength and direction of climate change.

In order to show the utilization of climate model for the investigation of future climate on a local scale the MPI-EMS-MR model air temperature data from Max-Plank Institute, Germany were used. The RCP 4.5 and RCP 8.5 scenarios were examined. The RCP 4.5 scenario presents average results and stabilization of the concentrations through the implementation of measures to limit greenhouse gas emissions. According to this scenario, greenhouse gas emissions will increase until 2040, after which the radiative forcing is expected to be 4.5 W/$m^2$ by 2100, which corresponds approximately to concentrations of about 650 ppm $CO_2$ eq. The RCP 8.5 scenario is the most pessimistic and implies a rapid increase in population, low technological development, without measures to reduce greenhouse gas emissions, increase poverty and, on the other hand, high energy consumption and increased emissions (van Vuuren et al, 2011). The radiative forcing will increase to 8.5 W/$m^2$ to the year 2100, which corresponds to concentrations of 1370 ppm in $CO_2$ eq.

The model data were downscaled to the region with coordinates: lon 21.56 - 23.44, lat 41.04 - 42.90, which includes parts of western Bulgaria and the eastern parts of Macedonia and Serbia. The area was chosen according to the availability of the observation data which are needed for the bias correction of model data. The research was done by simulations for two periods 2021-2050 and 2051-2080. The observation data for the period 1986-2015 was used as a reference period for bias correction of the model data. The temperature data are corrected for each month and year with the difference between observation data average for the period 1986-2015 and model data. The complete coincidence in the annual cycle of air temperature based on observation and model data is established after bias correction. The model data show negative deviations from the observation data only for the period May - June, while for the other months there are positive deviations (Tab. 1).

This example indicates the uncertainty of climate model data, the importance of observation data and the usefulness of bias corrections.

***Table 1. Monthly air temperature (°C) from the observation and MPI-EMS-MR model before and after bias correction for the period 1986-2015***

| | Jan | Feb | Mar | Apr | May | Jun | Jul | Aug | Sept | Oct | Nov | Dec |
|---|---|---|---|---|---|---|---|---|---|---|---|---|
| Before correction | 2.7 | 4.0 | 6.6 | 10.6 | 14.1 | 17.8 | 20.4 | 21.7 | 18.6 | 13.3 | 7.1 | 3.9 |
| After correction | -0.5 | 1.6 | 5.5 | 10.7 | 15.4 | 19.1 | 21.6 | 21.4 | 16.7 | 11.2 | 5.5 | 0.5 |
| Observation | -0.5 | 1.6 | 5.5 | 10.7 | 15.4 | 19.1 | 21.6 | 21.4 | 16.7 | 11.2 | 5.5 | 0.5 |

In order to examine the impact of the shadow economy on the national climate change policy, linear regression is applied. This method is chosen due to the following reasons: it is an efficient method in primary studies and it is simple and easy to use. The present paper is a first in its kind which studies the impact of the shadow economy and populism over the climate change policy. The results of the investigation could lead to the increasing of the scientific interest in this topic and also will put the attention to the shadow economy and populism as factors for risk and uncertainty of the climate change policy.

The linear regression analysis amongst the aforementioned factors, was performed using both the statistical software STATA (13.1) and the Microsoft office application - Excel. This could help to be calculated more accurately the correlations amongst factors and to be received the most reliable conclusions and results.

# 5 Results and discussion

## 5.1 Climate sensitivity analysis

According the RCP 4.5 scenario the changes in seasonal air temperature in the studied area are close to 0 and are statistically non-significant. The results from the investigation of tendencies in seasonal air temperature for two future 30-years periods (2021-2050 and 2051-2080) show well determined positive trend by RCP 8.5. (Tab. 2). The highest values are obtained for summer and autumn for the period 2051-2080.

***Table 2. The coefficients of linear trend (°C/ 10 years) of seasonal air temperature according to RCP 4.5 and RCP 8.5MPI-EMS-MR model data for the territory lon = 21.563 - 23.438, lat = 41.036 - 42.901***

| Scenarios | Period | Winter | Spring | Summer | Autumn |
|---|---|---|---|---|---|
| RCP 4.5 | 2021-2050 | 0.4 | 0.0 | 0.1 | 0.3 |
| | 2051-2080 | 0.4 | -0.3 | -0.1 | -0.3 |
| RCP 8.5 | 2021-2050 | **0.6** | 0.1 | **0.5** | 0.3 |
| | 2051-2080 | 0.1 | **0.5** | **1.4** | **0.7** |

** The values in Bold are statistically significant at p=0.05*

The increasing of air temperature is synchronous with the course of greenhouse gases. In 2006 the concentration of $CO_2$ reached a value of 403.3 ppm. According to the most unfavorable scenario RCP 8.5 (IPCC, 2013, AR5), the average annual temperatures for the Balkan Peninsula (including Bulgaria) are expected to increase by 0.5 to 0.75 °C for every 100 ppm $CO_2$ (eq) greenhouse gas concentrations. By the end of the century, the rise in temperatures would be from 5 to 7 °C, while the optimistic scenario RCP 2.6 showed a rise in temperature of no more than 2 °C (MOEW, 2014).

RCP scenarios show various trends of precipitation changes in Bulgaria. The average annual precipitation total in the country is expected to change by ±10% for the period 2016-2035, while at the end of the century the reduction will be between 10 and 20%. The most significant decrease is expected, according the RCP 8.5 scenario, at the end of the century - by 20-30%, and in Southeastern Bulgaria - by 30-40%(MOEW, 2014).

### 5.2 Shadow economy, populism and the national climate change policy

The regression analysis between greenhouse gas emissions intensity of energy consumption (dependent variable) and shadow economy (independent variable) confirms the main hypothesis of the present research that the share of the shadow economy has impact on the Greenhouse gas emissions intensity of energy consumption.

For the purposes of the research, the following variables are analyzed: 1) Independent variable: "Shadow economy" (ShadEcon) and 2) Dependent variable: "Greenhouse gas emissions intensity of energy consumption" (CIEC). The data for "Shadow economy" is taken from the research of Medina and Schneider (2018). Data for "Greenhouse gas emissions intensity of energy consumption" is taken from the EUROSTAT database (2020).

In order to examine if the correlation between the aforementioned variables could be performed through a linear function, the preliminary calculation tests with the statistical software STATA (13.1) were made. The detailed calculations could be found in Appendix I (Tabl. 6, 7, 8, 9, 10).

First of all, the relationship between two variables is necessary to be tested for heteroscedastcity or what is the variance of the observed cases. In this regard the Breush-Pagan test will be applied, as it is one of the most popular and reliable one to test homogeneity of variance in a linear regression. For that purpose, we need to formulate $H_0$ and $H_1$. $H_0$ will say that there is no homogeneity of variance of the variables (this means heteroscedasticity) and $H_1$ will say the opposite – there is homogeneity of variance (this means homoscedasticity). After performing the Breush-Pagan test, it is clear that Prob is 0.4836, which value is greater than $\alpha$ (0.05). Consequently we will confirm the alternative hypothesis ($H_1$) as the correct one, which says that there is a homogeneity of variance between the variables and the relationship among them could be presented by a linear function, Appendix I (Tabl. 6).

Secondly, the Ramsey Regression Equation Specification Error Test (RESET) is applied. The test shows that there are no omitted cases in the process of calculations. Consequently, it could be expected that no errors amongst the regressors will occur, Appendix I (Tabl.7).

Thirdly, variance inflation factor (VIF) test is applied in order to be determined the severity of multicollinearity. The test shows that VIF=1, which means that there is no correlation among the $j^{th}$ predictor and the remaining predictor variables, and hence the variance of (ShadEcon) is not inflated at all, Appendix I (Tabl. 8).

In addition, Akaike's information criterion and Bayesian information criterion test and Effect sizes for linear models are also made, Appendix I (Tabl. 9, 10).

Having in mind the presented information above, it is clear that the relationship between both variables (Shadow economy and Greenhouse gas emissions intensity of energy consumption) could be explained via linear regression. For that purpose, the authors will apply first the Microsoft Office application “Excel” and then the statistical software STATA (13.1) so as the results obtained and final conclusions to be confirmed two times and to be more reliable.

The independent variable is “Shadow economy” (ShadEcon) and the dependent variable is: “Greenhouse gas emissions intensity of energy consumption)”, (CIEC):

***Table 3. Linear Regression analysis with STATA***

| . | regress CIEC: | ShadEcon | | | |
|---|---|---|---|---|---|
| Source | SS | df | MS | Number of obs | = 16 |
| Model | 350.461737 | 1 | 350.461737 | F( 1, 14) | = 16.40 |
| Residual | 299.148416 | 14 | 21.367744 | Prob > F | = 0.0012 |
| | | | | R-squared | = 0.5395 |
| Total | 649.610153 | 15 | 43.3073435 | Adj R-squared | = 0.5066 |
| | | | | Root MSE | = 4.6225 |
| CIEC | Coef. | Std. Err. t | P>t | [95% Conf. | Interval] |
| ShadEcon | -.9293687 | .2294812 -4.05 | 0.001 | -1.421557 | -.4371805 |
| _cons | 134.2636 | 6.216855 21.60 | 0.000 | 120.9298 | 147.5975 |

The interpretation of the results obtained via STATA and Excel will be made at the same time. Detailed information about the Excel calculations could be found in Appendix II (Tabl.11).

As showed in Tabl. 11 (Appendix II), it is clear that there is a linear relationship amongst the studied variables as the Sign. F is 0,005175, which value is much lower than α (0,05). Consequently the results obtained could be interpreted (Excel).

Multiple R is 0,68, which means that there is a strong proportional relationship amongst the dependent and independent variables (Appendix II, Tabl. 11).

R Square is 0,539 STATA (Tabl.3) which means that 50% of the variations in the Greenhouse gas emissions intensity of energy consumption due to changes in the government policy, related to the control of Shadow economy.

Having in mind the results obtained by the linear regression model, the relationship between two variables – "Greenhouse gas emissions intensity of energy consumption" and "Shadow economy" could be presented via a linear function:

$Y = f(x_1 \ldots\ldots x_n) + \varepsilon$ or

$$Y = a + bx + \varepsilon$$

$$Y_{(CO2\ intensity\ of\ EC)} = a + b * x_{(shadow\ economy)} + \varepsilon,$$

where

a – constant (no interpretation is needed)

b – regression coefficient in front of the regressor

ε – residual

$$CO_2 \text{ intensity of EC} = 134{,}26 - 0{,}929 * \text{shadow economy} + \varepsilon$$

The equation above shows that if the shadow economy is decreased with 1%, then the "Greenhouse gas emissions intensity of energy consumption" will also decrease with 93 units. In this regard, the linear regression model allows shadow economy to be examined as a risk and uncertainty factor in the process of implementing the government policy for low carbon economy.

The model of Hammonds et al.,(1994) allows to measure the risk and uncertainty of the implementation of the main goals of the Bulgarian policy for climate change till 2030 for reducing the emissions of $CO_2$ in two scenarios (MOEW, 2012). In the first scenario, the Gg $CO_2$ eq have to be reduced to 60 943 and in the second scenario, when additional measures are introduced the Gg $CO_2$ eq have to be reduced to 52 642 until the 2030. If no measures for reducing the $CO_2$ emissions are introduced, their volume will be 66 360 Gg $CO_2$ eq. After the necessary calculations are done it is established that the risk of reaching the goals of the Bulgarian policy for climate change is 92%. The problem here is that factors such as force majeure, new global economic crises and change of the EU energy policy are not taken into account in these scenarios.

There is no reliable statistic information about the indicator “populism”, which information is needed for measuring the “climate populism” via a statistical model. The present study will try to establish a conceptual model for collecting statistical information for populism through the model of expert monitoring of the mass-media.

## 5.3 Study of climate change perception and understanding

The climate change perception and understanding are examined on the basis of the public opinion and in relation to the analysis of the existence of populism on the topic of climate change in Bulgaria mass media. This allows to determine the mass media effect on the topic of climate change and on the human perception. In order to study the public opinion the online survey was conducted. The survey was disseminated by e-mail and social networks and we have received the answers from 81 respondents. Although that the survey was sent by e-mail to the representatives of various sectors the most active were those working in the administration (including the state administration) and in the field of science and education (22% of completed surveys are from respectively of these two sectors, Tab.4). A relatively large number of completed surveys were received from the students and employees in sector "Services". Despite the significant number of respondents (81), the number of respondents from sectors such as industry, agriculture, NGOs is unsatisfactory.

***Table 4. Percentage of respondents who have taken a part in online survey***

| | Sectors | % of respondents |
|---|---|---|
| 1 | Industry | 4 |
| 2 | Services | 16 |
| 3 | Agriculture | 4 |
| 4 | Science and education | 22 |
| 5 | Administration | 22 |
| 6 | NGOs | 7 |
| 7 | Students | 17 |
| 8 | Other | 7 |

The first part of the survey aims to give the information about climate change perception. The results show that 55% of respondents believe there is a global threat as irreversible climate change. According to 27% the problem of climate change is highly significant and 21 % consider it as extremely significant (Fig. 1a, c).

Most of the participants in the survey understand that anthropogenic activity has an important role for climate change (Fig. 1, b) but 34% think that climate change is due to the equal impact of anthropogenic activity and natural factors. Only 1/3 part of respondents said they were directly affected by climate change (Fig 1 d).

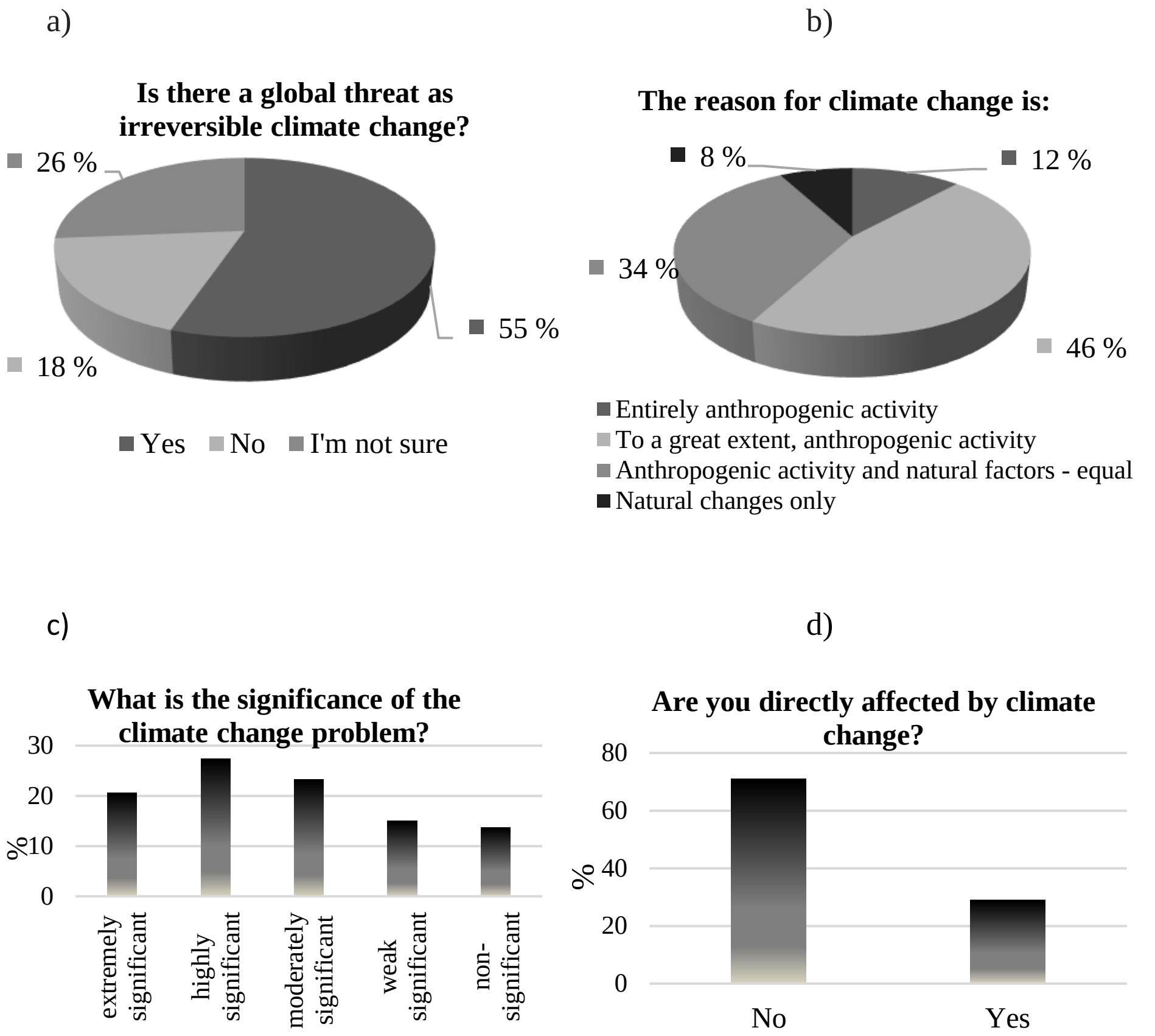

***Fig. 1 Climate change understanding and perception – answers from the survey (in percentage of the total number of respondents)***

The participants in the study indicate that climate change affects them by increasing of frequency of extreme meteorological and climate phenomena. The hailstorms, intense rainfall, floods, prolonged droughts destroy agricultural crops and also lead to the increasing of cost of property insurance. Some of the participants are affected by reducing water resources, shortening the ski season or exacerbating illnesses such as asthma and poly-allergy.

The public opinion about the Climate change issue completely responds to the inquiry of the Bulgarian mass media publications presented on Fig. 2., where 72% of the publications confirmed that the climate change is a fact. This is identical with the government position on the topic as well as with the positions of the international organizations.

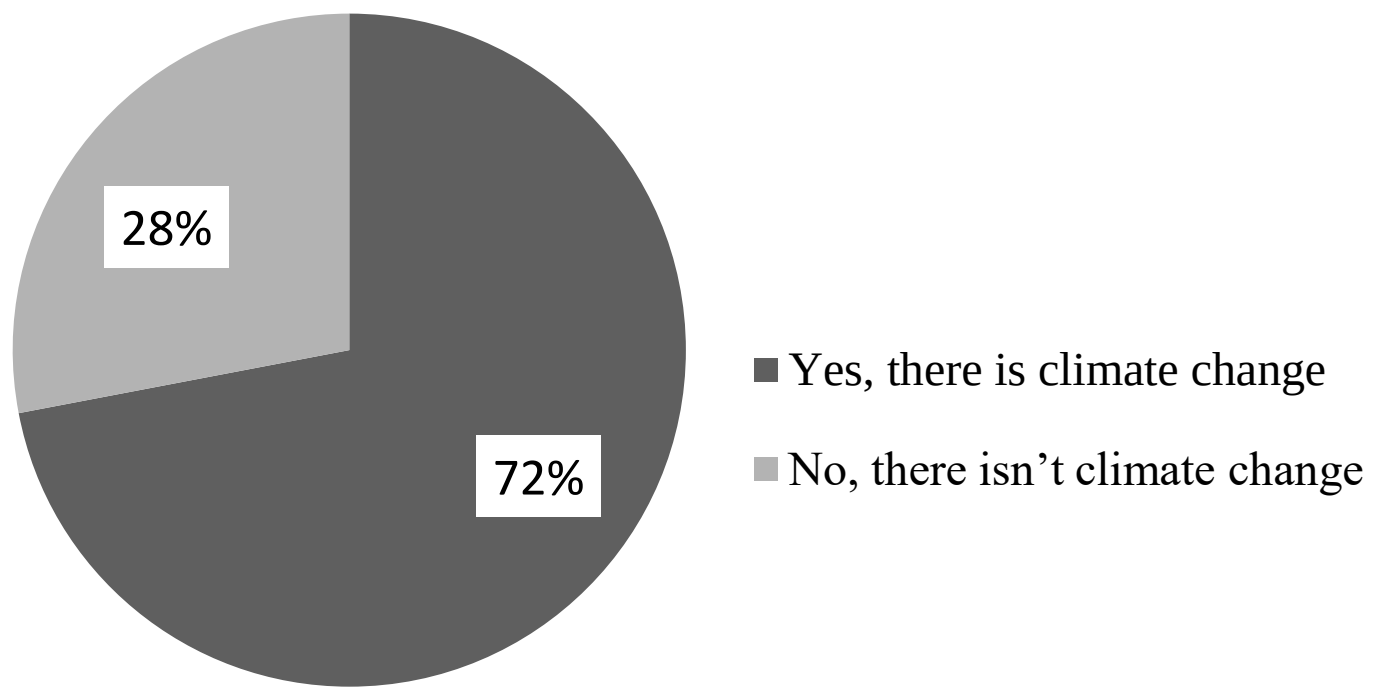

***Fig. 2. Global Warming: Is it Real? – analysis of 500 publications in the most popular Bulgarian media (2012-2017).***

The answers of the questionnaires and most of the publications in the mass media confirm the existence of inevitable effects of the climate change. However 44% of the publications inspire the conception that the society could not overcome the problem concerning climate change or it is too expensive for the society to overcome it (Fig. 3). This inspiration makes the Bulgarian society not to change the status quo. For example the energy sector in the country, which is the main polluter of the environment not to be reformed and etc. This is a populist statement.

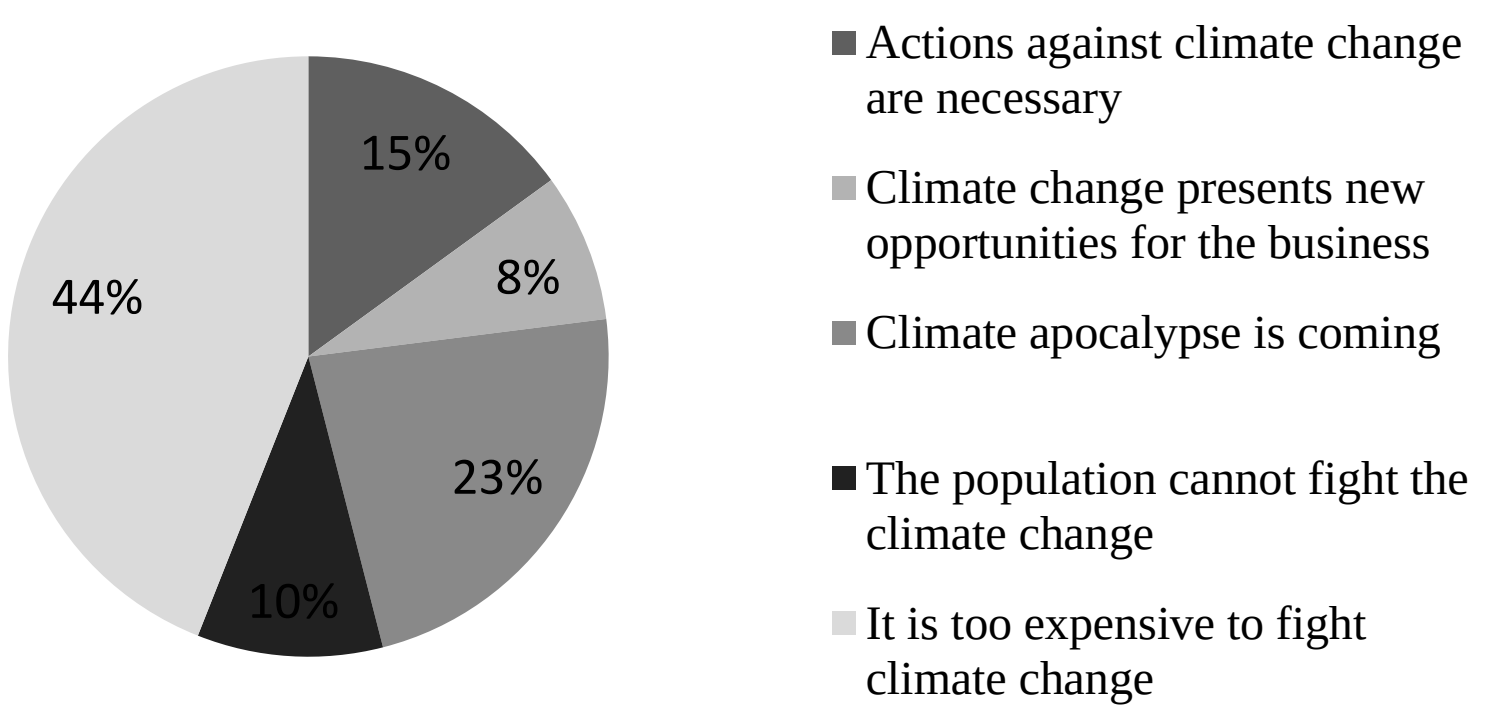

***Fig. 3. Climate Change Populism and the Media - Existence of extreme opinions in the mass media, concerning climate change - analysis of 500 publications in the most popular Bulgarian media (2012-2017)***

23% of the publications defend the thesis that if the economy is not stopped a climate apocalypse is coming. This is also a populist statement.

The last 23% of the publications, which are divided in two categories do not defend populist statements. According to the first category (15%), active policy measures against climate change should be taken. The second category (8%) point out that climate change presents new opportunities for the business, such as carbon finance, green energy and etc. This is completely identical with the government position on the topic as well as with the positions of the international organizations.

On the basis of the above analysis it can be concluded that 77% of the analyzed publications on climate change in the Bulgarian mass media for the period 2012-2017 are populist. This creates a risk of wrong impact over the social opinion on the topic about climate change. This also creates a risk for the government actions on the problem.

The second part of the online survey is dedicated to the problems related to the greenhouse emissions and the transition to low-carbon economy. It is extremely necessary to set targets and implement measures to reduce carbon emissions and mitigate climate change for 52% of the participants in the survey and 26% consider it as very necessary. Only for 1% it is not necessary at all and for 3% this is not very important.

This statement does not respond to the inquiry of the Bulgarian mass media publications, presented on Fig 3. Obviously the populist publications which are 77% of the total number of examined publications, do not impact over the public opinion, as 52% of the population think that active measures against climate change should be taken.

According to the respondents whole society should be involved in the process of tackling climate change. The responsible for solving climate change problems are also the government, all industrial enterprises and academic community (Tab. 5), only 4% of participants in the survey think that there is not a problem "climate change".

*Table 5.Who should be involved in solving climate change problem?*

| If there is a problem like climate change, who should be involved in solving it? | Percentage of respondents who chose the answer* |
|---|---|
| Government | 63 |
| Large industrial enterprises | 29 |
| All industrial enterprises, regardless of size and production | 50 |
| Agricultural companies | 18 |
| Academic community / science | 45 |
| Non-governmental organizations | 22 |
| Individuals | 16 |
| Society as a whole | 76 |
| There is no "climate change" problem | 4 |

* - more than one answer could be chosen

Despite of high number of respondents who think that the government should be involved in solving climate change problem only 32% of all participants at the survey think that the Bulgaria conducts a policy aimed at limiting greenhouse gas emissions in the atmosphere.

89% of respondents agree to pay a higher price for the energy from alternative sources if this would reduce the concentration of greenhouse gases and pollutants in the atmosphere and to limit climate change. The respondents were asked to choose how much their monthly expenses to be increased - up to 2 Bulgarian leva (BGN) [1], up to 5, 10 or up to 20 Bulgarian leva. The distribution of answers in the different categories is almost equal - 25, 18, 25 and 22% respectively.Similar results are provided by Kotceva and Mochurova (2012), who have established by questionnaires the willingness of Bulgarian population to pay more for energy from renewable energy sources.

According to the survey the main driving factors for the activities related to the decreasing of carbon emissions are air quality and healthy environment, requirements of the legislative documents and climate change (Fig. 4). Among the main obstacles to the work for carbon emission reduction are Lack of human resources, knowledge and experience, Lack of financial resources and Citizens' reluctance to change (Fig. 5).

[1] 1 BGN = 1.95583 Euro

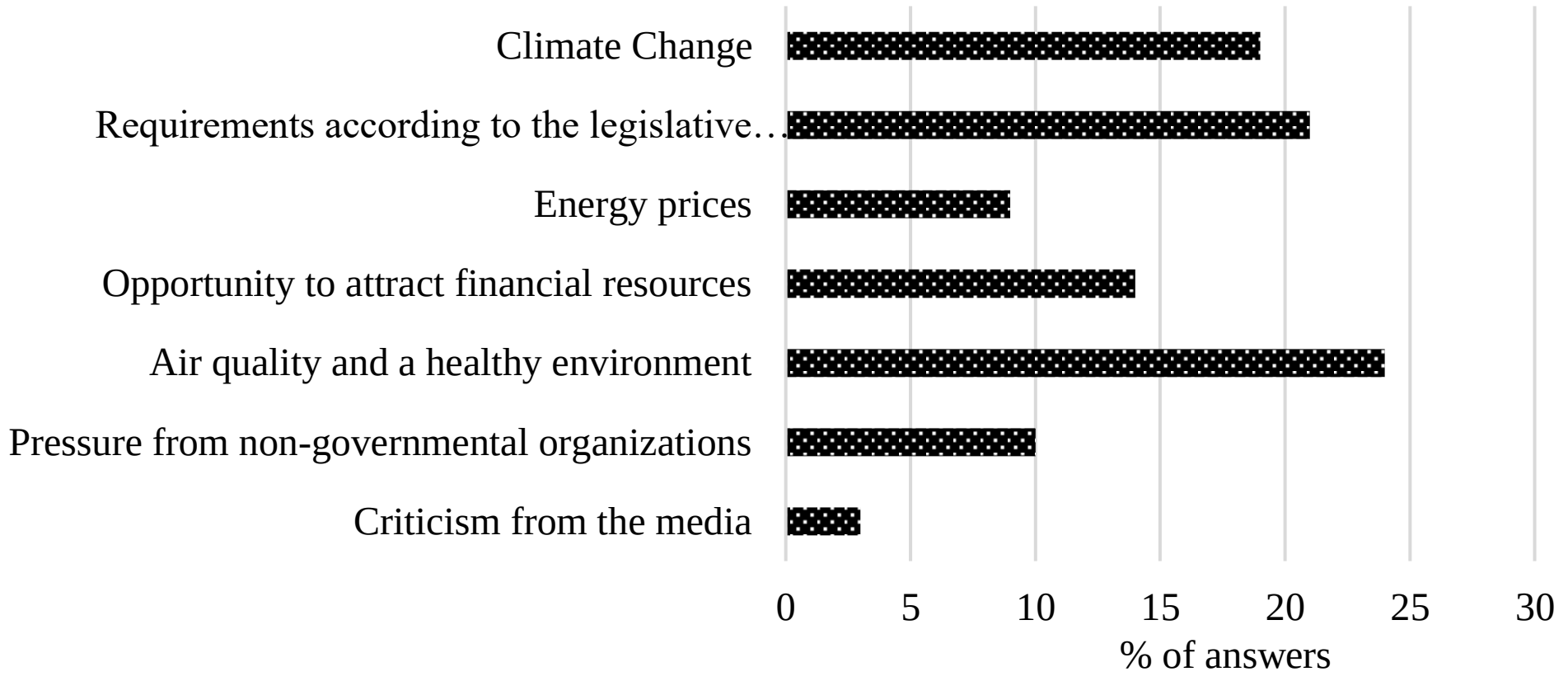

***Fig. 4. What are the driving factors for limiting carbon emissions activities?***

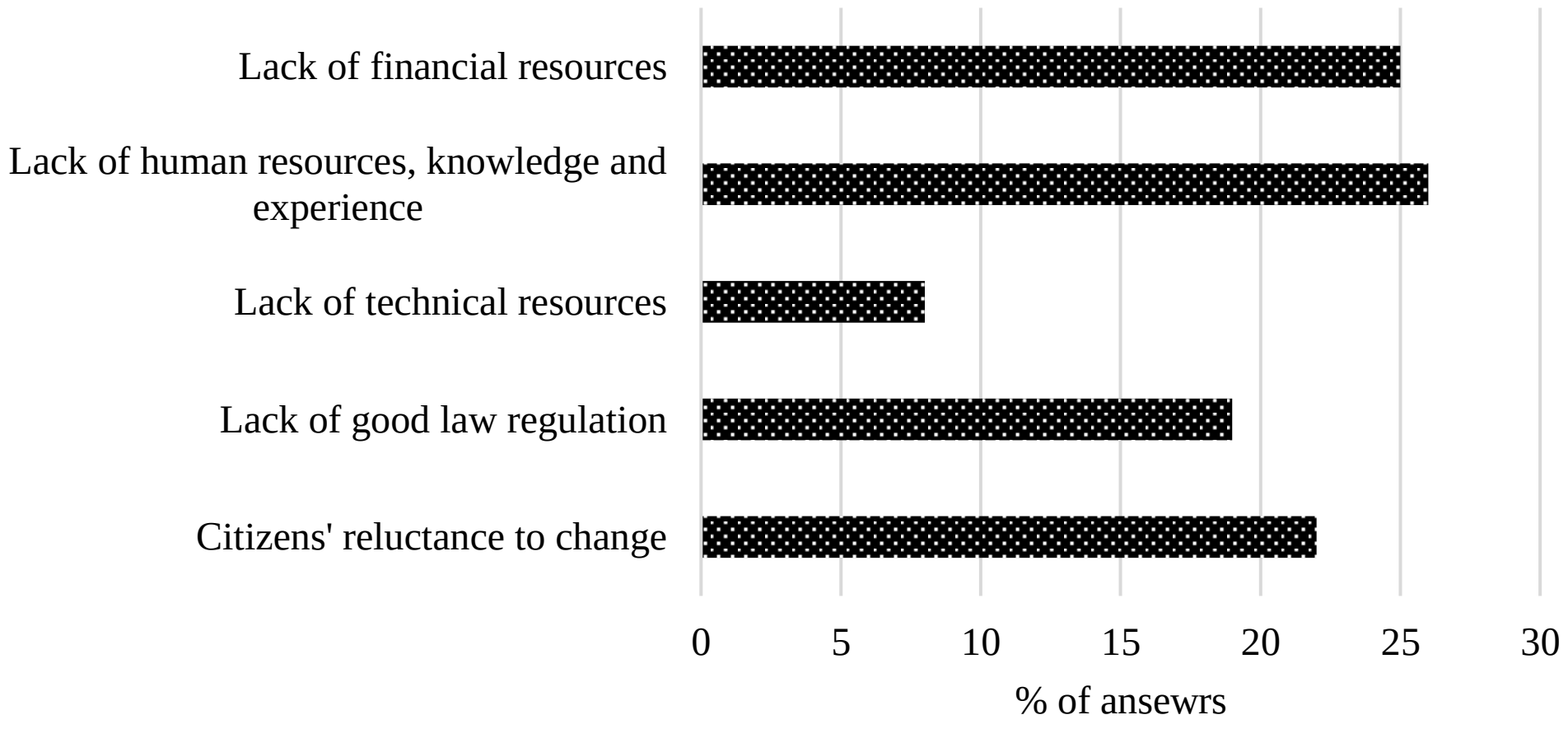

***Fig. 5. What are the main obstacles to the work for carbon emissions reduction?***

The above shown results have much in common with the information presented on Fig.3, according to which in 44% of the publications it is stated that it is too expensive for the society to fight climate change. From other side 25% of the respondents think that there is a lack of financial resources and over 20% is the citizens' reluctance to change. In 10% of the publications (Fig. 3) it is claimed that society cannot fight climate change. The results of the analysis allow to make the conclusion that mass media populism impacts over the public opinion.

## 6 Conclusions

The present analysis was made in the context of uncertainty and risk related to climate change and climate change policy. It is shown that the multicultural differences could be introduced as factors of risk and uncertainty in the realization of the climate change policy. The transition economy countries in the Balkans have their specifics. For all of them the percentage of shadow economy, corruption and populism is high. At the same time, some of them are EU member-states or EU Candidate Countries. The main problems, concerning the national policies for climate change of these countries could bring risk and uncertainty in the EU climate change policy. Bulgaria, Romania and Croatia have legal representatives in all of the EU institutions. The existence of populism could also be an obstacle the governments of these countries not to actively participate in the EU policymaking process and to put the EU policy for climate change at risk.

The regression analysis, as well as the qualitative and quantitative analyzes are executed by using statistical data only for Bulgaria, but it is supposed that the results obtained are also valid for the other countries from the Balkan Peninsula. As it is stated in part 2.1 of the present study, there are many similar features that unite Balkan countries. Such features/indicators are: share of shadow economy, populism, high energy and carbon intensity and etc. Consequently if some of these indicators are studied about one of the Balkan countries, the same assumptions could be made for the others. That is why we could confirm that conclusions made about the aforementioned indicators for Bulgaria could be valid also for the other Balkan countries.

The regression analysis indicates well determined correlation between the shadow economy and the Bulgarian climate change policy. A significant correlation between the indicators used for measuring the populism and the national climate change policy was not established. However, the link between them cannot be rejected and it could be proved by other indicators such as publications in the mass media, regarding extreme standpoints about climate change and their impact on public opinion. In depth quantitative and qualitative analyses have to be done in relation to these issue in the future.

## Appendix I.

### Preliminary tests for linear regression (STATA:13.1).

### *Table 6. Breusch-Pagan / Cook-Weisberg test for heteroskedasticity*

| |
|---|
| . estat hettest |
| Breusch-Pagan / Cook-Weisberg test for heteroskedasticity |
| Ho: Constant variance |
| Variables: fitted values of CIEC |
| chi2(1) = 0.49 |
| Prob > chi2 = 0.4836 |

### *Table 7. Ramsey Regression Equation Specification Error Test (RESET)*

| | | | | |
|---|---|---|---|---|
| . estat ovtest | | | | |
| Ramsey RESET test | using powers of the fitted | values | of | CIEC |
| Ho: model | has no omitted variables | | | |
| F(3, 11) = 2.28 | | | | |
| Prob > F = 0.1363 | | | | |

### *Table 8. Variance inflation factor (VIF) test*

| | | | |
|---|---|---|---|
| . | estat vif | | |
| | Variable | VIF | 1/VIF |
| ShadEcon | 1.00 | 1.000000 | |
| Mean VIF | 1.00 | | |

### *Table 9. Akaike's information criterion and Bayesian information criterion*

| |
|---|
| . estat ic |
| Akaike's information criterion and Bayesian information criterion |
| Model Obs ll(null) ll(model) df AIC BIC |
| . 16 -52.33329 -46.12983 2 96.25965 97.80483 |
| Note: N=Obs used in calculating BIC; see [R] BIC note |

### *Table 10. Effect sizes for linear models*

| | | | | |
|---|---|---|---|---|
| . estat esize | | | | |
| Effect sizes for linear | models | | | |
| Source | Eta-Squared | df | [95% Conf. | Interval] |
| Model | .5394955 | 1 | .1297887 | .7230792 |
| ShadEcon | .5394955 | 1 | .1297887 | .7230792 |

## Appendix II

## Regression analysis

### *Table 11. Regression analysis (Excel)*

SUMMARY OUTPUT

CIEC:ShadEcon

| *Regression Statistics* | |
|---|---|
| Multiple R | 0,681169169 |
| R Square | 0,463991436 |
| Adjusted R Square | 0,422760008 |
| Standard Error | 4,774329034 |
| Observations | 15 |

ANOVA

| | *df* | *SS* | *MS* | *F* | *Significance F* |
|---|---|---|---|---|---|
| Regression | 1 | 256,5111696 | 256,511 | 11,25334 | 0,0051747 |
| Residual | 13 | 296,3248304 | 22,7942 | | |
| Total | 14 | 552,836 | | | |

| | *Coefficients* | *Standard Error* | *t Stat* | *P-value* | *Lower 95%* |
|---|---|---|---|---|---|
| Intercept | 133,2810331 | 7,001707652 | 19,0355 | 7,08E-11 | 118,15476 |
| 35,3 | -0,8879045 | 0,264682642 | -3,3546 | 0,005175 | -1,4597166 |

| *Upper 95%* | *Lower 95,0%* | *Upper 95,0%* |
|---|---|---|
| 148,4073 | 118,1548 | 148,4073 |
| -0,31609 | -1,45972 | -0,31609 |